%% file: floquetRG3.tex
\documentclass[prx,twocolumn,notitlepage,superscriptaddress,noeprint]{revtex4-1}
\usepackage[left=1.5cm,right=1.5cm,top=1.8cm,bottom=1.85cm]{geometry}
\usepackage{amsmath}
\usepackage{amssymb}
\usepackage{graphicx}
\usepackage[caption=false]{subfig}
\usepackage{comment}

\newcommand{\ND}[1]{\hat{n}_{#1}}

\newcommand{\HamF}{\hat{\mathcal{H}}}

\DeclareMathOperator{\tr}{tr}

\input{macros.tex}

\begin{document}
\title{Flow Renormalization and Emergent Prethermal Regimes of Periodically-Driven Quantum Systems}
\author{Martin Claassen}
\affiliation{Department of Physics and Astronomy, University of Pennsylvania, Philadelphia, PA 19104, USA
}
\affiliation{Center for Computational Quantum Physics, Simons Foundation Flatiron Institute, New York, NY 10010 USA}

\begin{abstract}
We develop a flow renormalization approach for periodically-driven quantum systems, which reveals prethermal dynamical regimes and associated timescales via direct correspondence between real time and flow time behavior. 
In this formalism, the dynamical problem is recast in terms of coupling constants of the theory flowing towards an attractive fixed point that represents the thermal Floquet Hamiltonian at long times, while narrowly avoiding a series of unstable fixed points which determine distinct prethermal regimes at intermediate times. We study a class of relevant perturbations that trigger the onset of heating and thermalization, and demonstrate that the renormalization flow has an elegant representation in terms of a flow of matrix product operators. Our results permit microscopic calculations of the emergence of distinct dynamical regimes directly in the thermodynamic limit in an efficient manner, establishing a new computational tool for driven non-equilibrium systems.
\end{abstract}
\maketitle

\section{Introduction}	
One of the most fascinating aspects of non-equilibrium physics is that a macroscopic quantum system pushed out of equilibrium can exhibit markedly different dynamics when probed on different times scales. A prime emerging application is the design of materials properties out of equilibrium via strong optical driving fields. Here, broad tailored light pulses can ``dress'' the electronic, magnetic or lattice dynamics in what is commonly called Floquet engineering \cite{bukov15,oka18,rudner20}, to transiently stabilize new non-equilibrium states with novel or useful properties. Central to the feasibility of this approach, energy absorption from the driving field must be sufficiently suppressed such that the short and intermediate-time behavior of the driven system reflects a controlled modification of the electronic dynamics which can differ dramatically from the long-time thermal or steady-state behavior which is typically dominated by heating \cite{dalessio14,lazarides14}.

While time-resolved experiments in solid state and cold atomic systems are taking rapid strides in the transient control of single-particle and transport properties via light \cite{wang13,mahmood16,mciver20}, understanding the origin and microscopics of such prethermal dynamical regimes, predicting materials settings in which they can arise as well as devising theoretical descriptions of the associated time scales remain fundamental challenges. Important already in understanding deviations from quantized responses in the band engineering of Floquet topological insulators \cite{mciver20,sato19a,sato19b,nuske20}, such considerations are essential to extending the notion of Floquet control to interacting quantum systems. Here, the coherent induction of Floquet many-body phases naturally necessitates controlling both driving-induced modifications of the transient dynamics and the time scales of their validity, before heating inevitably takes over.

Theoretically, closed periodically-driven interacting quantum systems are expected at long times to heat to infinite temperature by virtue of the eigenstate thermalization hypothesis \cite{dalessio14,lazarides14}, in the absence of integrability or many-body localization \cite{dalessio13,ponte15,lazarides15}. However, heating and thermalization can in principle happen very slowly, leading to quasi-stationary prethermal plateaus at intermediate times. Most of our current numerical \cite{zeng17,machado19a,machado19b} and experimental \cite{singh19,rubio20,peng2021,kyprianidis21} understanding regards the limit of high-frequency driving, where the pump frequency $\Omega$ exceeds all local energy scales. Here, seminal recent works proved rigorous lower bounds for heating rates and the existence of prethermal dynamics for high-frequency driving \cite{abanin15,mori16,kuwahara16,ho17,abanin17a,abanin17b,else17,mori18,ho18} and weakly-kicked systems \cite{szabolcs18}, for which the short-time period-averaged dynamics can be described via effective time-independent Hamiltonians derived from perturbative Floquet-Magnus or Baker-Hausdorff expansions of the time evolution operator over a pump period, respectively. In these pictures, divergences at finite expansion order in inverse pump frequency signal the onset of heating \cite{mori16,ho17}; formally, an effective Hamiltonian can still be defined for the long-time dynamics, albeit at the expense of losing locality of interactions, and any local observables measured with respect to its eigenstates yield infinite-temperature expectation values \cite{dalessio14}.

In practice, a growing body of numerical \cite{claassen17,weidinger17,peronaci18,haldar18,gulden19,luitz19} and experimental \cite{rovny18} evidence points towards time scale separation and prethermal dynamics as a much broader phenomenon, that extends to resonant or low-frequency perturbations with emergent ergodic obstructions, separations of energy scales or protecting symmetries. Simultaneously, a sensible high-frequency driving limit is typically absent in real materials \cite{claassen16}, and the divergence of Floquet-Magnus expansions is known to be an insufficient signature of ergodic thermalization \cite{haga19}. Moreover, rigorous results for the onset of heating merely constitute worst-case bounds that could be significantly surpassed in  appropriately-chosen systems. A theoretical framework to describe such settings, to access emergent meta-stable dynamical regimes and predict the associated time scales is thus highly desirable, to extend light-induced Floquet engineering to strongly-correlated quantum systems and enable a host of new solid-state applications that range from the dynamical control of Mott insulators \cite{itin14,bukov15b,mentink15,kennes18b} or spin-orbital dynamics \cite{liu18,hejazi18,chaudhary19} to spin liquids \cite{claassen17} and fractional quantum Hall phases \cite{lee18,ghazaryan16}.

In this work, we present a flow renormalization framework for periodically-driven systems, for which a direct correspondence between real time and ``flow time'' behavior can be established. In this picture, the dynamical problem is recast in terms of couplings of the model flowing towards an attractive fixed point that represents the exact Floquet Hamiltonian, while narrowly avoiding a series of unstable fixed points which determine distinct dynamical regimes at intermediate times. This naturally elucidates emergent dynamical regimes and permits a straightforward extraction of the associated time scales. Crucially, we find that the relevant couplings at such prethermal fixed points are inherently non-perturbative, prohibiting any truncation of the flow equations in terms of local couplings. Instead, we demonstrate that the relevant dynamically-generated ''operator strings`` which are the root cause of the onset of heating have an elegant representation in terms exponentially-local infinite matrix product operators (MPOs). We show that this permits studying prethermal dynamical regimes, the loss of locality and the onset of thermalization directly in the thermodynamic limit in an efficient manner, establishing a new computational tool for non-equilibrium systems.

\section{Floquet quantum systems}
Consider an interacting quantum system subjected to a periodic drive with frequency $\Omega$. Its general Hamiltonian can be written as an expansion
\begin{align}
	\Ham(t) = \sum_{m} e^{i m \Omega t} \Ham_m  \label{eq:HamBare}
\end{align}
in harmonics $m$ with $\Ham_{-m} = \Ham_m^\dag$. A paradigmatic example is the canonical condensed matter setting of interacting electrons driven by light, with photons coupling to electrons via Peierls substitution. Here, a tight-binding description of electrons can be readily expressed in terms of harmonics $\Ham_m = \sum_{ij\sigma} \mathcal{J}_m(A |\mathbf{r}_{ij}|) e^{i m \arg \mathbf{r}_{ij}} g_{ij} \CD{i\sigma} \C{j\sigma}$, where $\mathcal{J}_m$ is the $m$-th Bessel function, $\mathbf{r}_{ij}$ are bond vectors, $g_{ij}$ denote hopping matrix elements between sites $i$ and $j$, and $A = a_0 e \mathcal{E}_0 / \hbar\Omega$ is the dimensionless strength of the electric field $\mathcal{E}_0$, written in terms of the lattice constant $a_0$ and electron charge $e$. For typical realizable drive strengths, $A \ll 1$; therefore, higher harmonics $|m| > 1$ can be neglected to treat a majority of experimental systems.

To describe the dynamics of the driven system, the time-evolution operator over a drive period $T = 2\pi / \Omega$
\begin{align}
	\hat{U}(T) = \hat{\mathcal{T}} e^{-i \int_{t_0}^{t_0+T} d\tau \Ham(\tau)} \equiv e^{-i \Ham_F T}   \label{eq:UT}
\end{align}
is the fundamental object of interest. It formally defines a Floquet Hamiltonian $\Ham_F$ that determines the period-averaged dynamics, up to a gauge freedom encoded in the arbitrary starting time $t_0$ \cite{bukov15}. Its eigenspectrum $e^{i \E_n T}$ determines Floquet quasi-energies $\E_n$, defined modulo $\Omega$ due to the absence of energy conservation in the driven system, and its many-body eigenstates take the form $\ket{\Psi_n(t)} = e^{-i\E_n t} \sum_m e^{im\Omega t} \ket{\Phi_{nm}}$. It is instructive to instead consider a frequency-domain representation
\begin{align}
	\HamF &= \sum_{m} \left[ m\Omega \otimes \ketbra{m}{m} + \sum_M \Ham_M \otimes \ketbra{m+M}{m} \right]  \label{eq:HamF}
\end{align}
defined in a product space of the original Hilbert space and the space of periodic functions \cite{sambe73}, with eigenstates $\ket{\phi_n}$ that obey $\ket{\Phi_{nm}} = \braket{m}{\phi_n}$. Note that the apparent Hilbert space expansion merely reflects a gauge redundancy of Floquet theory, as eigenstates with energy $\E_n$ and $\E_n + m\Omega$ identify with the same physical state $\forall m$. We henceforth denote conventional and frequency-domain operators using conventional $\Ham$ and calligraphic $\HamF$ notation, respectively. It is then straightforward to show [see Appendix \ref{app:TimeEvolutionOperator}] that the two-time evolution operator can be succinctly expressed as
\begin{align}
	\hat{U}(t,t') = \hat{\mathcal{T}} e^{-i \int_{t'}^{t} d\tau \Ham(\tau)}  &= \sum_m \bra{0} e^{-i \HamF (t-t')} \ket{m} e^{-im\Omega t'} \label{eq:SambePropagator}
\end{align}
While the sums over $m$ formally run over infinite Floquet indices, in practice convergence is guaranteed for truncations to $|m| \leq m_{\textrm{max}}$ with small $m_{\textrm{max}}$.

\begin{figure}[t]
	\centering
	\includegraphics[width=\columnwidth,trim=0.42cm 0cm 0cm 0cm]{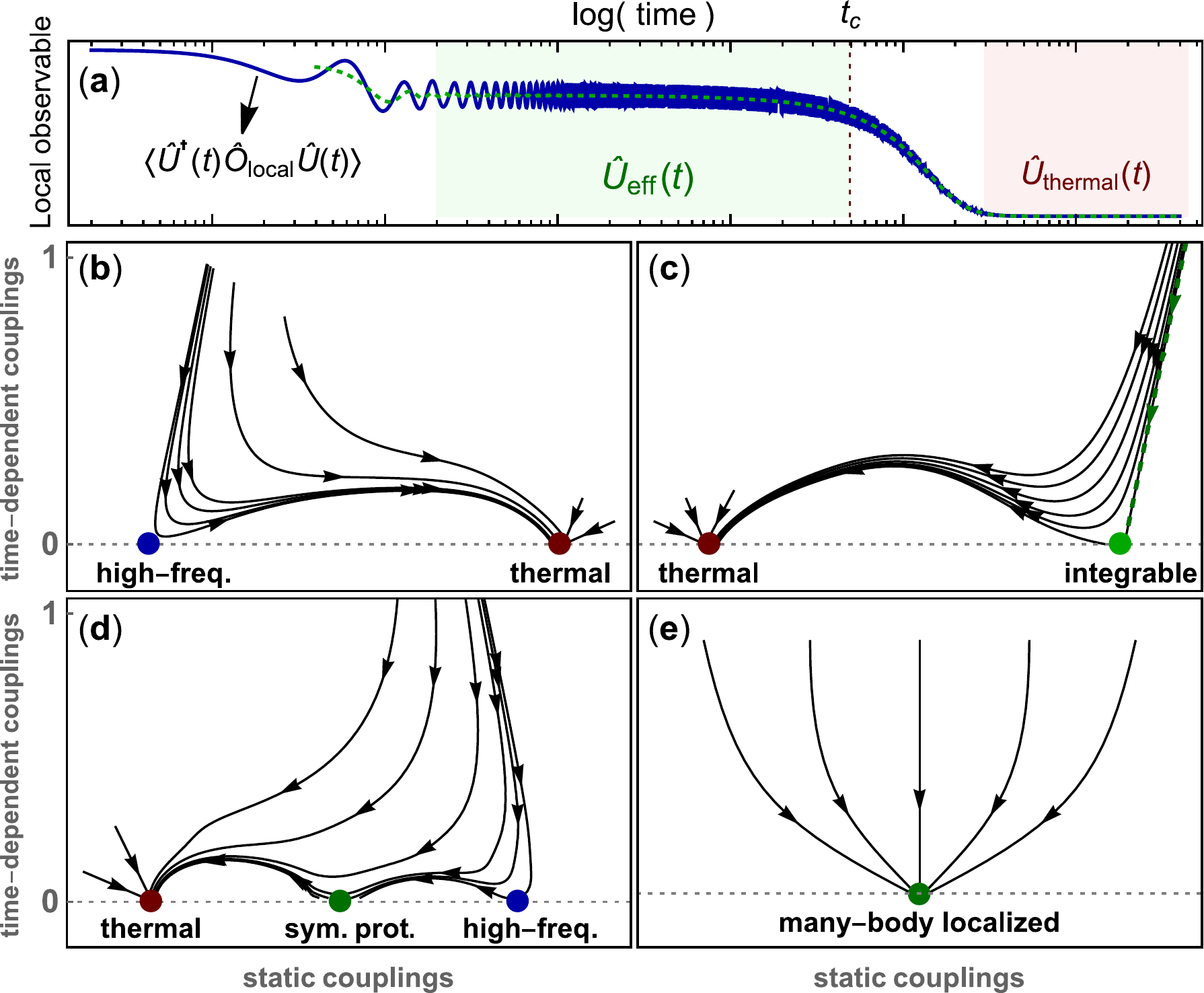}
	\caption{Time scales and fixed point landscape of Floquet flow renormalization. (a) Local observables in periodically-driven systems can exhibit quasi-stationary prethermal plateaus at intermediate times which are captured via an effective propagator. At long times, the system heats to a featureless infinite-temperature thermal state. (b) High-frequency driving and (c) Floquet integrability appear as unstable fixed points of Floquet flow renormalization, with the green dashed flow in (b) indicating the limiting case of an integrable theory. Relevant perturbations ultimately drive systems towards an attractive thermal fixed point.  Further scenarios include (d) the appearance of multiple unstable fixed points, as well as (e) many-body localization as a non-thermal attractive fixed point.}  \label{fig:schematics}
\end{figure}

Suppose now that a system is initially prepared in a quantum state $\ket{\Psi_0}$ and subjected to periodic driving. Physically, as closed ergodic driven quantum systems are expected to heat to infinite temperature, 
the long-time behavior of any \textit{local} physical observable should approach its infinite-temperature expectation value, independent of the initial state $\ket{\Psi_0}$ \cite{dalessio14}. Information about the latter is  ``scrambled'' over the macroscopic system \cite{nandkishore15,dalessio16} and becomes inaccessible to local probes. Instead, the Floquet Hamiltonian $\Ham_F$ that determines the long-time behavior is a non-local operator and its true eigenstates, for purposes of local measurements, are essentially random vectors, with $\hat{U}(T)$ exhibiting properties of random matrices drawn from a circular Gaussian ensemble \cite{dalessio14}.

A meaningful notion of Floquet engineering nevertheless survives if heating is sufficiently suppressed such that the intermediate-time dynamics differ drastically from the long-time thermal behavior. Fig. \ref{fig:schematics}(a) illustrates this scenario schematically for the time evolution of a generic local observable. After an initial ramping on of the external drive, its period-averaged expectation value saturates to a metastable plateau at intermediate times. While formally still governed by the exact Floquet Hamiltonian, this prethermal regime is instead well-described via an \textit{effective} time-evolution operator $\hat{U}_{\textrm{eff}}(T)$ and its associated effective Floquet Hamiltonian. The latter remains a \textit{local} operator and for judicious choices of the driving protocol reflects a controlled modification of the intermediate-time dynamics of interest -- the central target of Floquet engineering in correlated systems. Conversely, the effective description breaks down at long times; instead, the approach to the thermal regime is governed by a time-evolution operator with properties of a random matrix that reflects the infinite-temperature behavior.

\section{Floquet Flow Renormalization}
We are now in a position to formulate the renormalization problem for Floquet systems. Parameterize $\Ham(t)$ by a new scale $\lambda$, with a scale dependence of both time-independent $\Ham_0(\lambda)$ and drive $\Ham_{m \neq 0}(\lambda)$ parts such that $\Ham(\lambda = 0,t)$ corresponds to the original ``bare'' theory. Consider now a generic flow in $\lambda$ that proceeds to progressively eliminate the periodic time dependence $\Ham_{m\neq 0}(\lambda)$ due to the external drive, via moving into a rotating frame. This procedure implies attractive fixed points $\Ham(\lambda \to \infty, t) \equiv \Ham_0(\lambda \to \infty)$ of the flow which are time-independent [$\Ham_{m\neq 0}(\lambda \to \infty) \to 0$] and formally determine the ``full'' (non-local) Floquet Hamiltonian.

As the absence of energy conservation precludes a preferential energetic ordering and hence a recipe for the progressive elimination of degrees of freedom of the system, a straightforward choice is to demand that the flow should be unitary in the sense of conserving the eigenspectrum $e^{i\E_n T}$ of $\hat{U}(T) \to \hat{U}(\lambda,T)$, or equivalently $\HamF(\lambda)$. Na\"ively, if such a flow is performed exactly, $\Ham_0(\lambda \to \infty)$ at these fixed points will thus capture the exact renormalized period-averaged dynamics that represent the Floquet-engineered behavior of interest, whereas any knowledge of the residual micro motion within individual drive periods is encoded in the transformation to the rotating frame.

Unitary flows have a rich history in physics, and \textit{a priori} permit an infinite set of choices for the generator of the transformation that can be chosen to elucidate the physical properties of interest. Originally introduced by Glazek, Wilson and Wegner \cite{wilson93,wegner94} as a differential version of numerical discrete diagonalization methods, convergence properties were studied for numerous transformation choices \cite{henon74,white02,monthus16,savitz17}. Recent applications include studies of emergent conserved quantities in many-body localization \cite{monthus16,pekker16,thomson18,varma19,yu19} and derivations of spin and t-J models in Mott insulators \cite{reischl04,yang10}. In Floquet systems, truncated flow equations using Wegner-Wilson generators were introduced as a starting point for perturbative high-frequency expansions of the Floquet Hamiltonian \cite{verdeny13}, to study integrals of motion in Floquet many-body localization \cite{thomson20}, and an alternate transformation was recently studied with much improved convergence at lower drive frequencies \cite{vogl19} over Magnus expansions \cite{bukov15}.

Central to this work, rather than seeking optimal convergence to the Floquet Hamiltonian, we instead inquire whether a Floquet flow renormalization scheme may be formulated for which we may attribute physical meaning to the flow itself. Indeed, in generic quantum systems that obey the eigenstate thermalization hypothesis (ETH), the system heats and the true Floquet Hamiltonian $\Ham_F$ becomes a random matrix \cite{dalessio14} that represents the infinite-temperature behavior at long times, hence losing any notion of locality to become a physically ill-defined quantity. Thus, instead of seeking improved but in principle uncontrolled approximations of $\Ham_F$, we will show that a preferential choice of flow permits inferring distinct dynamical regimes and associated time scale separations directly from the fixed-point structure at finite $\lambda$.

Conveniently, three simple physical requirements suffice to uniquely implement this procedure. (1) First, a natural prerequisite for such an identification is that the flow scale $\lambda \sim [\textrm{time}]$ enters as an ``RG time'' with units of time. This stands in contrast to the Wegner-Wilson framework \cite{wilson93,wegner94} where $\lambda \sim [\textrm{time}^2]$.  (2) As the flow should strive to eliminate the periodic time-dependence, the generator of the transformation at any $\lambda$ should depend only on the time-dependent couplings $\Ham_{m \neq 0}(\lambda)$. Finally, (3) demand that the flow does not generate higher harmonics while progressively eliminating lower harmonics. Aside from being physically sensible, this requirement is essential to feasibly study regimes of strong external driving that lie beyond the purview of perturbative expansions, and without having to resort to artificial truncations of the renormalized harmonics.

Minimal flow equations that satisfy these requirements can be uniquely implemented for the frequency-domain Hamiltonian of Eq. (\ref{eq:HamF}) as
\begin{align}
	\frac{\partial \HamF(\lambda)}{\partial \lambda} =  \sum_{\substack{m \\ M>0}} \left[ \left(\Ham_M(\lambda) \otimes \ketbra{m+M}{m} - \hc \right),~ \HamF(\lambda) \right]
\end{align}
and preserve the quasi-energy spectrum of $\HamF$ (and by extension $\hat{U}(T)$) by virtue of unitarity. Reexpressed in terms of the harmonic components of renormalized Hamiltonian, one readily arrives at a coupled set of operator equations of motion that describe the renormalization of individual harmonic components:
\begin{align}
	\frac{\partial}{\partial \lambda} \Ham_m(\lambda) &= -m \Omega \Ham_m(\lambda) + \sum_{m'=0}^{m-1} \left[ \Ham_{m-m'}(\lambda),~ \Ham_{m'}(\lambda) \right] \notag\\
		&+ 2 \sum_{m' \geq m+1} \left[ \Ham_{m'}(\lambda),~ \Ham^\dag_{m'-m}(\lambda) \right]  \label{eq:flowGeneric}
\end{align}
These flow equations constitute the central starting point of this paper. All fixed points obey that the time-dependent components must vanish $\Ham_{m \neq 0} \to 0$. The flow is unitary and hence preserves the quasi-energy spectrum of the Floquet time-evolution operator (\ref{eq:SambePropagator}). One can see by inspection that no higher harmonics are generated under the flow if the time-dependence of the bare Hamiltonian is bounded at a maximum harmonic. Notably, the flow generator thus differs fundamentally from Wegner-Wilson prescriptions \cite{wilson93,wegner94,verdeny13,thomson20}, as the generator itself must not follow from a commutator with the flowing Hamiltonian.

For many practical applications, higher harmonics of the bare driven Hamiltonian are negligible and it suffices to consider the simplest scenario of a harmonically-driven system
\begin{align}
	\Ham(t) = \Ham_0(\lambda) + \left( \Ham_1^{\vphantom{\dag}}(\lambda) e^{i\Omega t} + \hc \right)
\end{align}
In this case, the associated flow equations for the time-independent Hamiltonian and harmonic driving term take a particularly simple form
\begin{align}
	\frac{\partial}{\partial \lambda} \Ham_0(\lambda) &= 2 \left[ \Ham_1(\lambda),~ \Ham_1^\dag(\lambda) \right] \notag\\
	\frac{\partial}{\partial \lambda} \Ham_1(\lambda) &= -\Omega~ \Ham_1(\lambda) - \left[ \Ham_0(\lambda),~ \Ham_1(\lambda) \right]  \label{eq:flowHarmonic}
\end{align}
Notably, invariance of the Floquet quasi energies entails a conservation of
\begin{align}
	\frac{1}{2} \frac{\partial}{\partial\lambda} \lVert \Ham_0(\lambda) \rVert^2 + \frac{\partial}{\partial\lambda} \lVert \Ham_1(\lambda) \rVert^2 + 2 \Omega \lVert \Ham_1(\lambda) \rVert^2 = 0
\end{align}
as an invariant of the flow.

Physical insight into the behavior of the flow may be readily gleaned from considering a generic model that evades thermalization to infinite temperature at long times. Particularly simple examples include high-frequency driving in the limit $\Omega \to \infty$, or integrable models such as free fermions driven by light or Ising chains in oscillating magnetic fields. In such cases, the bare Hamiltonian (\ref{eq:HamBare}) must flow towards an attractive fixed point, at which the time dependence vanishes. Importantly, the renormalized time-independent Hamiltonian at the fixed point $\Ham_0(
\lambda \to \infty)$ \textit{remains local} and describes a drive-induced modification of the electronic or magnetic couplings.

Suppose now that a perturbation is added to the original Hamiltonian $\Ham(\lambda=0, t) \to \Ham(\lambda=0,t) + \delta \Ham'$ which restores ergodicity to the system and ensures that the system heats. Energy is now absorbed from the pump, such that at long times any thermalized local observable reflects infinite-temperature behavior independent of the initial state. This implies that the corresponding true Floquet Hamiltonian effectively becomes a random matrix, which requires the existence of a distinct ``thermal'' attractive fixed point described by a highly \textit{non-local} theory.

The perturbation $\Ham'$ hence acts as a \textit{relevant perturbation}, rendering the original ``prethermal'' fixed point unstable [Fig. \ref{fig:schematics}(b)-(e)]. If the introduced scale $\delta$ is sufficiently small, the renormalization flow will first progressively eliminate the time dependence of the bare Hamiltonian and approach the original (local) fixed point arbitrarily closely (realizing an almost time-independent renormalized Hamiltonian $\Ham(\lambda)$ at intermediate flow times), before diverging at flow times $\lambda \sim \delta^{-1}$ to ultimately flow towards the attractive $T \to \infty$ ``thermal'' fixed point.

\begin{figure}[t]
	\centering
	\includegraphics[width=\columnwidth]{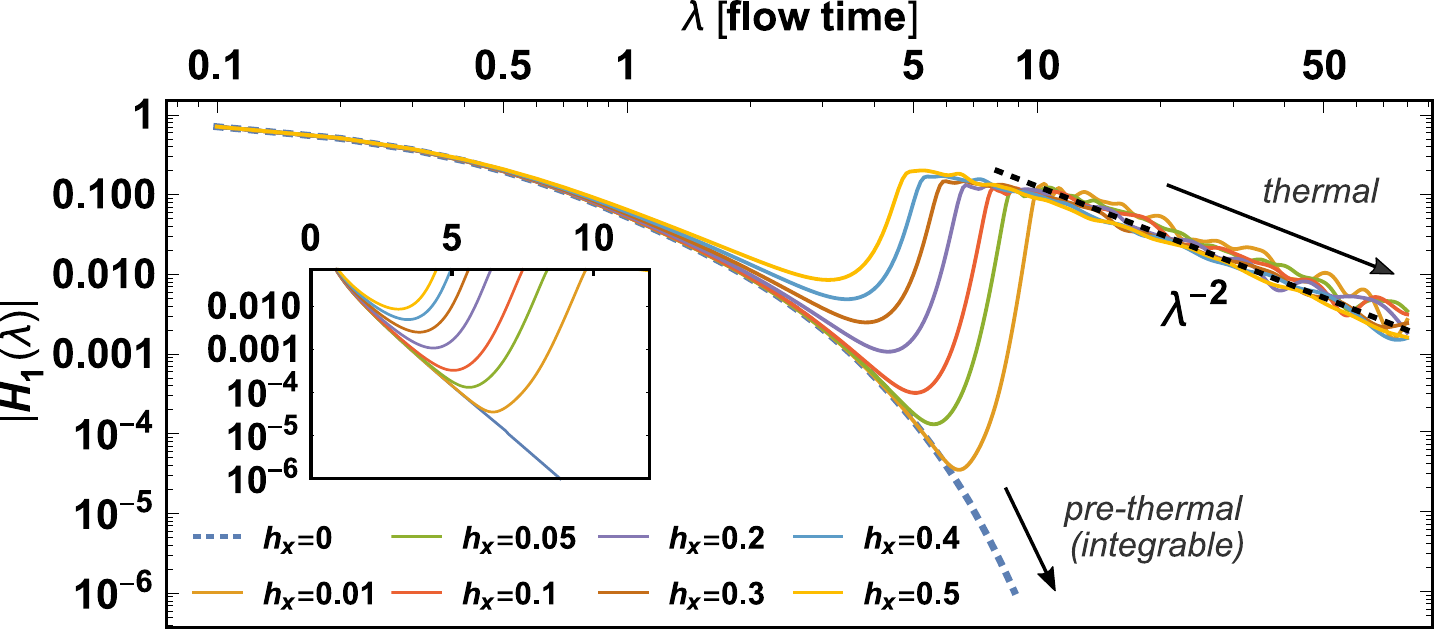}
	\caption{Flow renormalization of a driven transverse Ising model with $h_z = 0.7 + 0.5 \cos(3.5 t)$ in units of $J=1$, as a function of integrability-breaking parallel fields $h_x$, depicted for the magnitude $|\Ham_1|$ of the time-dependent couplings. While the model at small $\lambda$ flows towards an integrable prethermal fixed point, weak $h_x \neq 0$ integrability breaking acts as a relevant perturbation and pushes the system towards a thermal fixed point independent of initial parameters.}  \label{fig:isingFlow}
\end{figure}

To illustrate these considerations, first consider the simple case of a one-dimensional harmonically-driven transverse-field Ising chain, with a weak integrability-breaking parallel field $h_x$:
\begin{align}
	\Ham(t) = \sum_{i} \left[ -J\hspace{0.05cm} \hat{S}_{i+1}^x \hat{S}_i^x + \left( h_z + A \cos(\Omega t) \right) \hat{S}_i^z + h_x \hat{S}_i^x \right]  \label{eq:isingModel}
\end{align}
In the absence of $h_x$, the Hamiltonian is integrable via Jordan-Wigner transformation. The model hence flows towards a ``prethermal'' fixed point described by a static renormalized $\Ham_0^{\textrm{pre}}$ of free Jordan-Wigner fermions, with the harmonic time dependence $\Ham_1(\lambda)$ eliminated exponentially in flow time $\lambda$ upon approaching the fixed point. Fig. \ref{fig:isingFlow} (dashed line) evaluates this scenario via exact diagonalization for an $L=14$ site chain and depicts the flow of the total magnitude of time-dependent couplings, defined as the (intensive) operator norm $\lVert \Ham_1(\lambda) \rVert^2 \equiv \tr\{ \Ham_1^\dag \Ham_1 \} / (L \tr\{ \mathbf{1} \})$ of the harmonic drive as a function of flow time $\lambda$.

\begin{figure*}[t]
	\centering
	\includegraphics[width=10cm,trim=0 0.5cm 0 0]{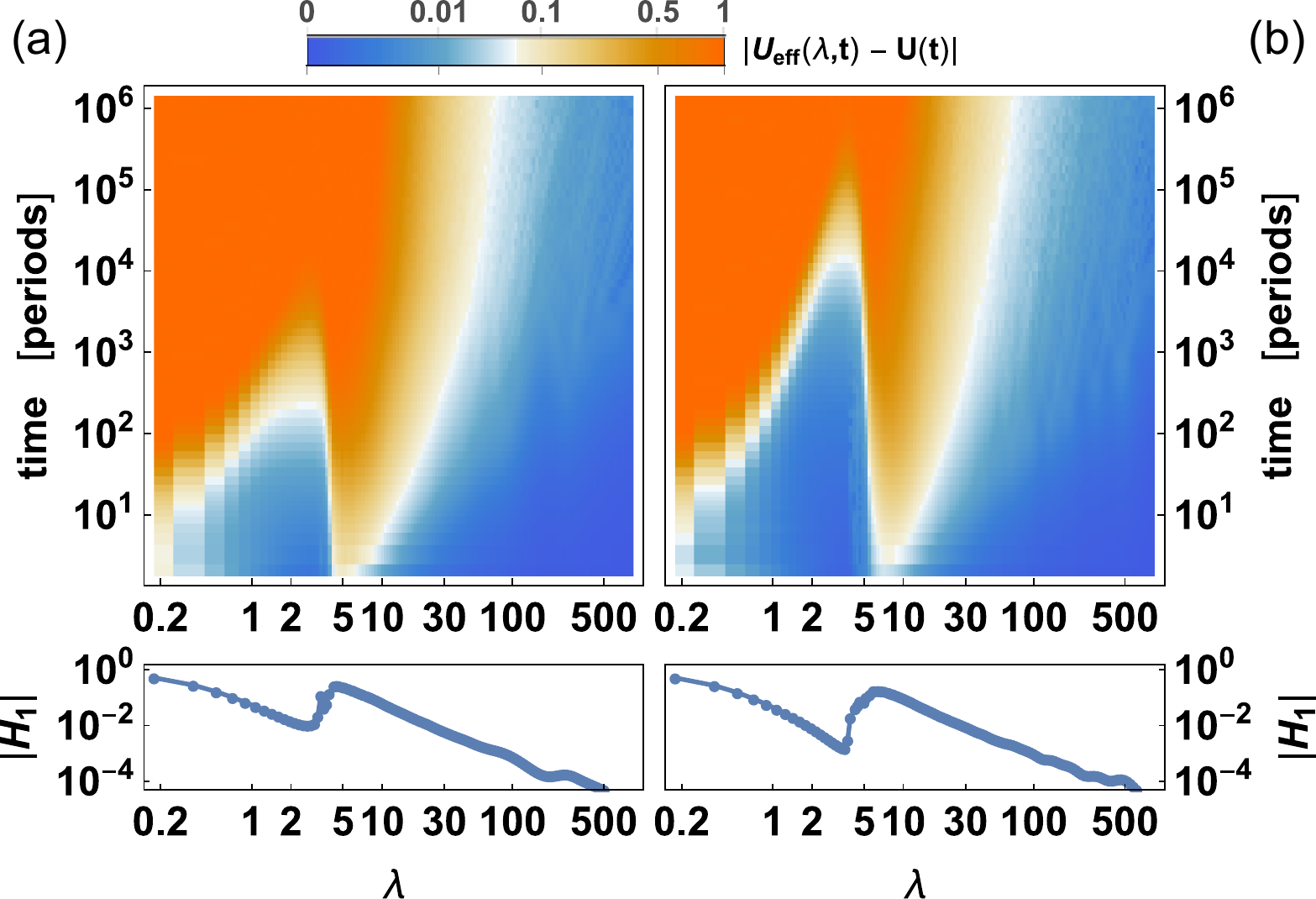}
	\caption{Emergence of distinct dynamical regimes in a driven 16-site transverse Ising model $\Omega = 3.5$ and parallel fields (a) $h_x=0.5$, (b) $h_x=0.2$ (units of $J=1$). Top panels depict the time-dependent deviation of the exact time evolution operator from effective propagators $U_{\textrm{eff}}(\lambda,t) = \exp(i \Ham_0(\lambda) t)$ defined via solely the static renormalized Hamiltonian at finite $\lambda$, with the flow of time-dependent components depicted below. Short- and intermediate-time dynamics are captured via effective local Hamiltonians dictated by the pre-thermal fixed point, whereas a thermal fixed point at large $\lambda$ describes the long-time dynamics.}  \label{fig:isingTimeEvolution}
\end{figure*}

A weak parallel field $h_x$ now breaks integrability and causes the flow to narrowly avoid the prethermal fixed point as a function of $h_x$ [Fig. \ref{fig:isingFlow}]. The magnitude of time-dependent couplings $\lVert \Ham_1(\lambda) \rVert$ becomes minimal in the vicinity of the prethermal fixed point, however diverges at longer flow times. This behavior coincides with a loss of locality of $\Ham_0(\lambda)$, and the model instead flows towards the attractive infinite-temperature thermal fixed point. Information about the initial prethermal dynamics is scrambled, which is reflected in the scaling collapse of flows of the time-dependent couplings independent of $h_x$ as $\lambda \to \infty$ [Fig. \ref{fig:isingFlow}].

Naturally, proximity to integrability does not exhaust the possible fixed point landscape. First, prethermal high-frequency driving \cite{abanin15,mori16,kuwahara16,ho17,abanin17a,abanin17b,mori18,ho18} canonically appears as an unstable fixed point [Fig. \ref{fig:schematics}(b)]. Here, if the drive frequency $\Omega$ exceeds local energy scales of the system, the flow of time-dependent couplings $\Ham_1(\lambda)$ [Eq. (\ref{eq:flowHarmonic})] is dominated at early flow times via exponential decay $\sim e^{-\Omega \lambda}$, and a prethermal fixed point can be approached exponentially closely as a function of increasing frequency [Fig. \ref{fig:schematics}(b)], before relevant perturbations redirect the flow towards the thermal fixed point. Furthermore, many-body localization manifests as an alternate attractive non-thermal fixed point of the flow [Fig. \ref{fig:schematics}(e)] that describes the long-time behavior of disordered interacting driven systems.

Crucially and more generally, any ergodic bottleneck to heating and thermalization maps onto new unstable fixed points of the flow, permitting for instance the capture and study of obstructions to energy absorption due to symmetries or separations of energy scales at low frequency driving. This constitutes a fundamental advantage of the flow renormalization scheme: as no energy scale is treated preferentially, the fixed point structure reflects solely the fundamental dynamical transition between locality and loss of locality of the effective Floquet Hamiltonian on different time scales; the flow renormalization formalism hence provides an unbiased tool to detect prethermal regimes and separations (and extraction) of time scales in generic driven systems.

An interesting consequence is the possibility to realize multiple prethermal dynamical regimes in a driven system [Fig. \ref{fig:schematics}(d)]. A simple physical realization is a system that can be separated into two weakly-coupled components, where one component is integrable in isolation and the second component exhibits a prethermal plateau. Here, in addition to the prethermal and thermal fixed points discussed above, a third fixed point naturally arises to describe the intermediate-time regime during which some degrees of freedom already exhibit thermalizing dynamics whereas a metastable plateau persists for a subset of the system.

\section{Real-time dynamics and heating}
To see how the fixed point landscape is reflected in the real-time dynamics, suppose now one stops the flow at flow time $\lambda$ near a prethermal fixed point. As the time-dependent couplings $\Ham_1(\lambda)$ flow to a minimal non-zero value and the renormalized Hamiltonian $\Ham(\lambda,t)$ is therefore almost stationary, one can ask how well the real-time evolution is captured by dropping the residual time dependence of the renormalized Hamiltonian and instead using the effective time evolution operator
\begin{align}
	\hat{U}_{\textrm{eff}}(\lambda, t, t') = e^{-i \Ham_0(\lambda) ~(t-t')}  \label{eq:Ueff}
\end{align}
to model the intermediate-time behavior. A simple metric is given by the operator distance with respect to the exact time-evolution operator (\ref{eq:SambePropagator}) after $n$ periods $T = 2\pi/\Omega$
\begin{align}
	d^2(\lambda, nT) &= \int_0^T \frac{dt_0}{2T} \left\lVert \hat{U}_{\textrm{eff}}(\lambda, t_0 + nT, t_0) - \hat{U}(t_0 + nT, t_0) \right\lVert^2 \notag\\
		&= 1 - \textrm{Re} \tr\left\{ e^{i\HamF_0(\lambda) nT} e^{-i\HamF(\lambda) nT} \rho_{\infty} \right\} / \tr\{ \rho_{\infty} \}  \label{eq:distanceMetric}
\end{align}
with the starting time $t_0$ averaged over a single pump period to integrate out the gauge freedom of the Floquet description. Here, $\lVert \hat{O} \rVert^2 \equiv \tr\{ \hat{O}^\dag \hat{O} \} / (L \tr\{ \mathbf{1} \})$ is again defined as an intensive operator norm for operators $\hat{O}$ and system size $L$, $\rho_{\infty} = \mathbf{1} \otimes \ketbra{0}{0}$ denotes a Floquet infinite-temperature ensemble in Sambe space, and $\HamF_0(\lambda) = \sum_m [ \Ham_0(\lambda) + m\Omega ] \otimes \ketbra{m}{m}$. The operator distance $d(\lambda, nT)$ is zero if $\hat{U}_{\textrm{eff}}(\lambda, nT)$ faithfully represents the time-evolution operator at time $nT$, whereas it saturates to $d(\lambda, t \to \infty) \to 1$ as the exact and effective time evolution operators dephase with respect to each other. It is straightforward to show that the distance is bounded from above by a linear rate [see Appendix \ref{app:DistanceMetric}]
\begin{align}
	d(\lambda, nT) ~\leq~ nT \left\lVert \sum_{M>0} \Ham_M(\lambda) \right\rVert  \label{eq:propagatorDistanceBound}
\end{align}
given precisely by the total magnitude of the time-dependent couplings as depicted in Fig. \ref{fig:isingFlow}. For the harmonic drive studied above, this naturally defines a time scale
\begin{align}
	t_{\textrm{eff}}(\lambda) = \left\lVert \Ham_1(\lambda) \right\rVert^{-1}
\end{align}
for the applicability of the effective theory $\hat{U}_{\textrm{eff}}(\lambda,t,t')$ [Eq. (\ref{eq:Ueff})].

Crucially, as $\lVert \Ham_1(\lambda) \rVert$ (or $\lVert \sum_{M>0} \Ham_{M} \rVert$ for general drives) becomes minimal near the prethermal fixed point with $\frac{\partial}{\partial \lambda} \lVert \Ham_1(\lambda) \rVert_{\lambda \to \lambda_{\textrm{pre}}} = 0$, this defines a prethermal time scale
\begin{align}
    t_{\textrm{c}} = t_{\textrm{eff}}(\lambda_\textrm{pre})
\end{align}
that determines the time regime of validity of the description of the dynamics in terms of the prethermal fixed point, with an effective static Floquet Hamiltonian $\Ham_0(\lambda_{\textrm{pre}})$. This establishes a straightforward but powerful new procedure to determine dynamical regimes and associated time scales for generic driven quantum systems: here, instead of directly solving for their real-time dynamics, a typically highly-challenging task, one can instead find a controlled non-perturbative solution of the full flow renormalization equation (\ref{eq:flowGeneric}) (to be discussed in the sections that follow), which reveals distinct dynamical regimes as an avoided fixed point of the flow for small $\lambda$. The corresponding prethermal time scale follows directly from the minimum of time-dependent couplings near this fixed point, with the corresponding dynamically-engineered Floquet Hamiltonian given by the renormalized static component $\Ham_0(\lambda_{\textrm{pre}})$. The latter then effectively governs the real-time evolution starting from any initial state, and remains valid for the duration of the prethermal plateau. Notably, this stands in contrast to bounds established for the high-frequency driving regime \cite{abanin15,mori16,kuwahara16,ho17,abanin17a,abanin17b,else17,mori18,ho18}; here, the flow of $\Ham_1(\lambda)$ permits a microscopic determination of the relevant time scales for generic systems and drive protocols, with all energy scales and ergodic obstructions of the system treated on equal footing.

To exemplify this behavior, Fig. \ref{fig:isingTimeEvolution} depicts $d(\lambda, t)$ as a function of flow time and real time on hand of the tranverse-field Ising model with broken integrability, evaluated exactly for a 16-site chain with periodic boundary conditions for two choices of the parallel field. As the flow approaches the prethermal fixed point near which the time-dependent couplings $\Ham_1(\lambda)$ become minimal, a long-lived prethermal regime emerges. Its dynamics are effectively captured by the renormalized time-independent Hamiltonian $\Ham_0(\lambda_{\textrm{pre}})$. This Hamiltonian is local and exhibits a controlled dynamical modifications of the magnetic interactions, however remains valid only up to a parametrically-large time $t_{\textrm{c}}$ that depends on the minimal magnitude of $\lVert \Ham_1(\lambda) \rVert$ near the prethermal fixed point, hence the strength of the parallel field $h_x$. Conversely, as the Hamiltonian flows towards the thermal fixed point at long flow times $\lambda \to \infty$, the time-independent component $\Ham_0(\lambda \to \infty)$ approaches the exact Floquet Hamiltonian and correctly describes the infinite-temperature dynamics at long real times, however at the expense of loss of locality of the description as expected.

This separation of time scales has immediate ramifications for pump-induced heating. Suppose that the system of interest initially starts from an initial state $\ket{\Psi}$. In a driven system, an appropriate measure of heating is the period-averaged absorbed energy $E(t) = \expect{ \hat{U}(t_0,t_0+t) \Ham_0(\lambda) \hat{U}(t_0+t,t_0) }$ expressed in terms of the renormalized Hamiltonian in the rotating frame. As the time-dependent couplings become minimal near the prethermal fixed point $\lambda_{\textrm{pre}}$, the residual time-dependence $\Ham_1(\lambda_{\textrm{pre}})$ can be readily treated as a perturbation. A linear-response calculation of the energy absorption rate \cite{abanin15} yields after integrating out the Floquet gauge freedom
\begin{align}
	\frac{\partial E}{\partial t} &= 2 \textrm{Im}~ e^{i\Omega t} \frac{\partial}{\partial t} \int_0^t d\tau \braOPket{\Psi}{ \left[ \Ham_1(\tau), \Ham_1^\dag(0) \right] }{\Psi} e^{i\Omega (\tau-t)}
\end{align}
with $\Ham_1(t) = e^{i \Ham_0(\lambda_\textrm{pre}) t} \Ham_1(\lambda_\textrm{pre}) e^{-i \Ham_0(\lambda_\textrm{pre}) t}$.

Importantly, this approach is applicable since the renormalized driving term $\Ham_1(\lambda)$ is weak, even if the bare system is strongly driven. By comparing the interaction-picture equation of motion $i\partial_t \left( e^{i\Omega t} \Ham_1(\lambda,t) \right) = -\Omega \Ham_1(\lambda,t) - [\Ham_0(\lambda), \Ham_1(\lambda,t)]$ with the exact flow equations (\ref{eq:flowHarmonic}) and iterating, one finds $e^{i\Omega t} \Ham_1(\lambda, t) = \Ham_1(\lambda + it) + \mathcal{O}(\lVert \Ham_1(\lambda) \rVert^3)$, which entails that the heating rate at short times follows from an analytic continuation of flow time $\lambda \to \lambda + it$
\begin{align}
	\frac{\partial E}{\partial t} &= 2 \textrm{Im}~ e^{i\Omega t} \frac{\partial}{\partial t} e^{-i\Omega t} \int_0^t d\tau \braOPket{\Psi}{ \left[ \Ham_1(\lambda + i\tau), \Ham_1^\dag(\lambda) \right] }{\Psi}
\end{align}
establishing a formal correspondence between real time and flow time behavior near the prethermal fixed point. This can be understood from noting that the flow equations near the fixed point can be linearized, with the evolution of the flow in $\lambda$ analogous to imaginary-time evolution. These equations in principle permit a direct determination of heating rates for a given initial state $\ket{\Psi}$. More broadly however, energy absorption is generically slow in the prethermal regime for arbitrary initial states, as the suppression of $\Ham_1$ near the prethermal fixed point entails that the time-independent component $\Ham_0(\lambda)$ becomes an almost-conserved operator, quantified via: 
\begin{align}
	C(t) &= \sqrt{ \int\displaylimits_0^T \frac{dt_0}{T} \left\lVert \hat{U}(t_0,t_0+t) \Ham_0(\lambda) \hat{U}(t_0+t,t_0) - \Ham_0(\lambda) \right\rVert^2 } \notag\\
	&\leq 2 t \cdot \left\lVert ( \partial_\lambda + \Omega) \Ham_1(\lambda) \right\rVert
\end{align}
at $t = nT$ integer multiples of the pump period, with straightforward generalizations to drives with higher harmonics [see Appendix \ref{app:ConservedH0}].

\section{Relevant Perturbations}
Central to this work, instead of directly tackling the real-time evolution of driven quantum systems, the flow renormalization scheme permits to determine the effective dynamics and associated time scales via looking for the fixed-point behavior of the flow equations. If the latter is known, this constitutes a key advantage, as time-domain studies of strongly-interacting quantum systems far from equilibrium typically pose profound challenges due to entanglement growth and a severe dynamical sign problem. Conversely, as the Hilbert space of the renormalized operators scales exponentially in complexity with system size, a practical implementation of the flow necessitates expressing the flowing Hamiltonian in terms of a subset of flowing couplings which must encode the essence of the dynamical regimes of interest. A fundamental question therefore concerns determining which are these relevant operators that are necessary to correctly describe the fixed point dynamics and the onset of heating.

To evaluate the flow renormalization beyond exact solutions for small systems, a simple but too na\"ive strategy is to expand the flowing Hamiltonian $\Ham(\lambda, t)$ in a constrained basis of few-body operators, and truncate the exact flow equations (\ref{eq:flowGeneric}) in this basis. The resulting flow equations for a vastly-reduced set of coupling coefficients can then be readily integrated. This approximation is commonplace in studies of the Wegner-Wilson flow equations, for instance in studies of higher-order corrections to magnetic exchange coefficients in Mott insulators \cite{reischl04,yang10}, and has been previously applied to continuous unitary transformations for Floquet systems to efficiently approximate the full Floquet Hamiltonian beyond the high-frequency limit \cite{verdeny13,vogl19,thomson20}.

To illustrate the inability of this approach to discern distinct dynamical regimes, consider the flow renormalization of a driven one-dimensional $J$, $J'$ Heisenberg model
\begin{align}
	\Ham(t) = \left[ J + A \cos(\Omega t) \right] \sum_{\left<ij\right>} \hat{S}_i \cdot \hat{S}_j + J' \sum_{\left<\left< ij \right>\right>} \hat{S}_i \cdot \hat{S}_j
\end{align}
with SU(2) spin rotation symmetry, where $A/J = 0.5$, $J'/J = 0.6$ and $\Omega=3.5J$ in units of $J=1$. Subjected to the flow equations of Eq. \ref{eq:flowHarmonic}, the renormalized Hamiltonian develops progressively higher-order spin interactions, which can be readily expressed in terms of sums of SU(2)-invariant operator strings. By virtue of time-reversal symmetry, $\Ham_0(\lambda) = \sum_{ij} J_{ij}(\lambda) \hat{S}_i \cdot \hat{S}_j + \sum_{ijkl} K_{ijkl} (\hat{S}_i \cdot \hat{S}_j) (\hat{S}_k \cdot \hat{S}_l) + \dots$ can be expanded solely in terms of strings of products of Heisenberg exchange couplings, whereas the flow of the harmonic drive $\Ham_1(\lambda)$ will additionally generate strings of SU(2)-symmetric scalar spin chirality terms $\sum_{ijk} \chi_{ijk}(\lambda) \hat{S}_i \cdot ( \hat{S}_j \times \hat{S}_k ) + \dots$.

\begin{figure}[t]
	\centering
	\includegraphics[width=\columnwidth]{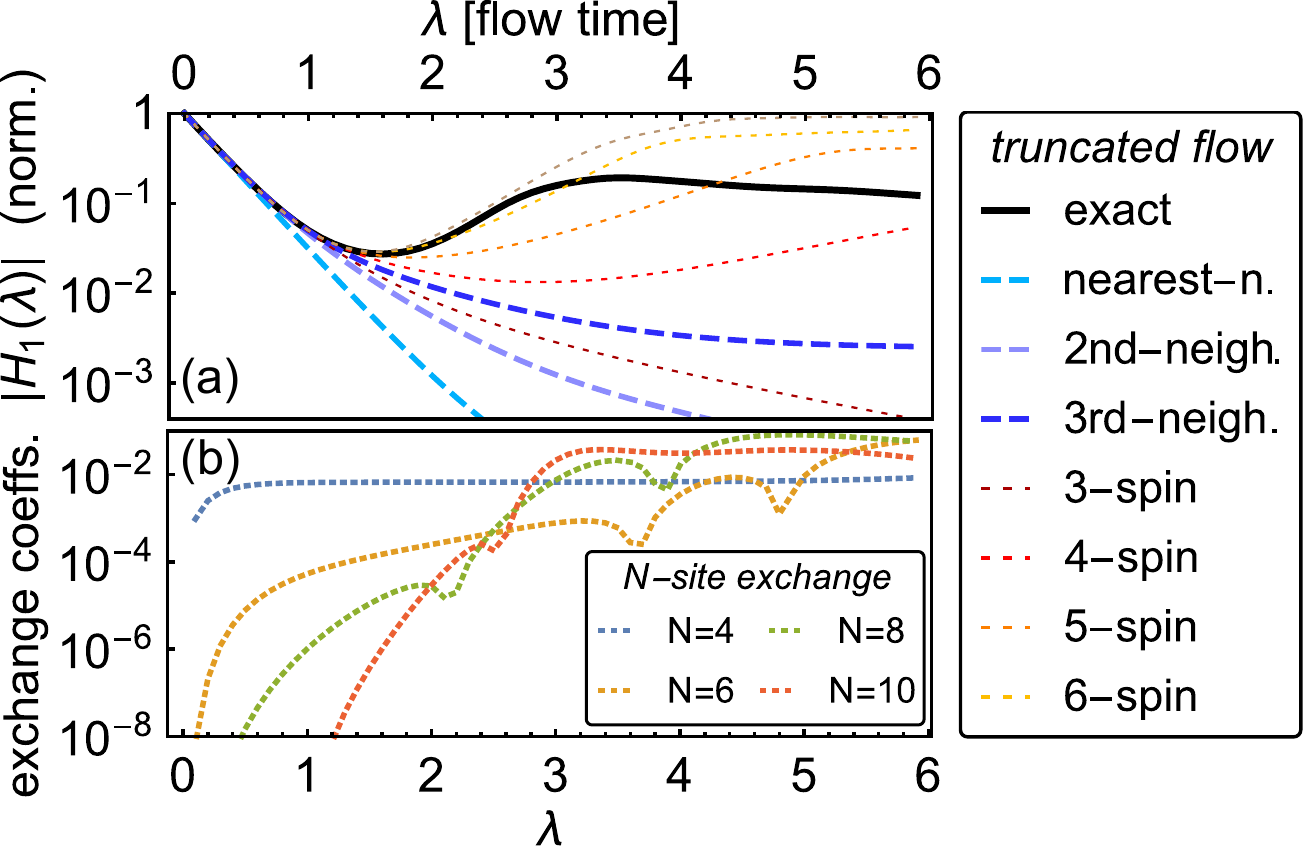}
	\caption{Truncated flow renormalization for a $J_1, J_2$ Heisenberg model with $J_1 = J + 0.5 \cos(\Omega t)$, $J_2 = 0.6 J$ with $\Omega = 3.5J$ in units of $J=1$. (a) shows the flow of the total magnitude of time-dependent couplings as a function of the truncation scheme, either by constraining the range of interactions (up to 3rd-nearest-neighbor) or by truncating the maximum $N$-spin interactions (with arbitrary range) generated under the flow (up to 6-spin interactions). Both truncation schemes fail to resolve the prethermal minimum until the truncation threshold exceeds the system size. (b) depicts the flow of individual couplings for four representative examples of 4-spin to 10-spin exchange.} \label{fig:truncatedFlow}
\end{figure}

Fig. \ref{fig:truncatedFlow}(a) compares the exact renormalization flow with results for approximate flow equations, depicted by example of the magnitude of time-dependent couplings $\lVert \Ham_1(\lambda) \rVert$ for a short chain of 12 spins. While such a system is too small to exhibit thermalization of individual states at long times,  the collective flow of all eigenstates of the renormalized $\Ham_0(\lambda)$ nevertheless exhibits a fixed point structure that mirrors a macroscopic system. The approximate flow equations are truncated either to a finite range of spin interactions (up to third-nearest-neighbor exchange) or to at most $N$-spin exchange interactions with no constraints on locality. The former approximation reproduces the initial renormalization of $\Ham(\lambda,t)$ at short flow times $\lambda$, however fails to captures the avoided prethermal fixed point. No minimum of the magnitude of flowing time-dependent couplings $\lVert \Ham_1(\lambda) \rVert$ develops, and the flow instead converges to an artificial fixed point at long flow times. This behavior is expected, as relevant perturbations near the prethermal fixed point must trigger a loss of locality of the Hamiltonian that signifies the onset of thermalization, which is precluded if longer-ranged couplings are discarded. Notably, no prethermal time scale and associated effective theory can be deduced.

Importantly, the situation remains unchanged if no constraints on locality are made and the flow equations are instead truncated to permit generating non-local exchange interactions, however limited to $N$-spin term [Fig. \ref{fig:truncatedFlow}(a)]. Remarkably, only after accounting for \textit{long-ranged and high-order multispin interactions} does the approximate flow reproduce the prethermal minimum of $\lVert \Ham_1(\lambda) \rVert$ hence permit determining the real-time behavior.

The root cause lies in the rapid growth of individual multispin exchange interactions of $\Ham_0(\lambda)$ near the prethermal fixed point, as depicted in Fig. \ref{fig:truncatedFlow}(b) for the coupling coefficients of four operator strings $\hat{O}_N = (\hat{S}_i \cdot \hat{S}_{i+1}) (\hat{S}_{i+2} \cdot \hat{S}_{i+3}) \cdots (\hat{S}_{i+N-2} \cdot \hat{S}_{i+N-1})$ of the exact flow, up to $N=10$. Treating such interactions on equal footing is therefore essential to capture the prethermal minimum and time scales, however precludes an approximative treatment of the flow equations in terms of straightforward truncations to finite-length operator strings.

\section{Matrix Product Operator Floquet Renormalization Flow}
Matrix product operators (MPOs) represent an ideal remedy of this conundrum. MPOs express generic many-body operators in terms of products of operator-valued matrices which are composed of operators that act strictly locally at a single site. For instance, the Ising model with transverse and parallel fields [Eq. (\ref{eq:isingModel})] has a compact MPO representation
\begin{align}
	\hat{W}_i &= \left[\begin{array}{ccc}
		\hspace{0.15cm} \hat{\mathbf{1}} \hspace{0.15cm} & J \hat{S}_i^x  & h_x \hat{S}_i^x + h_z \hat{S}_i^z \\
		0 & 0 & \hat{S}_i^x \\
		0 & 0 & \hat{\mathbf{1}}
	\end{array}\right]~~ \begin{array}{ll} \mathbf{l} &= \left[ 1,~ 0,~ 0 \right]^\top \\
		\mathbf{r} &= \left[ 0,~ 0,~ 1 \right]
	\end{array} \\
	\Ham &= \mathbf{l} \cdots \hat{W}_{i} \cdot \hat{W}_{i+1} \cdot \hat{W}_{i+2} \cdots \mathbf{r} \notag\\
		&= J \sum_{\left<ij\right>} \hat{S}_i^x \hat{S}_j^x + \sum_i \left( h_x \hat{S}_i^x + h_z \hat{S}_i^z \right)
\end{align}
given by a rank-4 tensor (operator-valued matrix) $\hat{W}$ with bond dimension three and boundary vectors $\mathbf{l}$, $\mathbf{r}$. For translation-invariant systems, $\hat{W}$ is identical at each site and $\Ham$ can therefore be extended to the thermodynamic limit as an infinite MPO (iMPO) via pushing the boundary to infinity. Notably, while individual eigenstates as well as time evolutions of quantum states can exhibit a high degree of entanglement and hence incur costly matrix product state representations with high bond dimensions, the representation of operators and operator dynamics remains comparatively poorly understood, but recently garnered attention as a tool to address scrambling and the spreading of local operators under unitary time evolution \cite{bohrdt16,xu18,hemery19}. Notably, MPOs were recently demonstrated to permit an efficient implementation of the Wegner-Wilson flow equations for many-body localized Hamiltonians \cite{yu19} that exploits the locality of the $l$-bit representation.

Central to this paper, efficient representations of operators as MPOs with low bond dimension are neither constrained to strictly-local operators nor limited to few-body interactions. To see this, consider an MPO with bond dimension three of the form
\begin{align}
	\hat{W}_i &= \left[\begin{array}{ccc}
		\hspace{0.05cm} \hat{\mathbf{1}} \hspace{0.05cm} & J \hat{S}_i^x  & 0 \\
		0 & e^{-1/\zeta} \hat{A}_i & \hat{S}_i^x \\
		0 & 0 & \hat{\mathbf{1}}
	\end{array}\right]  \label{eq:exponentialMPO}
\end{align}
where $\hat{A}_i = a_0 \hat{\mathbf{1}}_i + a_x \hat{S}_i^x + a_y \hat{S}_i^y + a_z \hat{S}_i^z$ is a generic normed local operator. The corresponding Hamiltonian
\begin{align}
	\Ham = J \sum_{i < j} e^{-\frac{|i-j-1|}{\zeta}} \hat{S}_i^x \left( \prod_{k=i+1}^{j-1} \hat{A}_k \right) \hat{S}_j^x
\end{align}
is exponentially local, however compactly encodes an infinite sum of higher-order $N$-spin operators $\hat{S}_i^x \hat{A}_{i+1} \hat{A}_{i+2} \cdots \hat{A}_{i+N-2} \hat{A}_{i+N-1} \hat{S}_{i+N}^x$ that are exponentially suppressed as a function of their support.

\begin{figure}[t]
	\centering
	\includegraphics[width=\columnwidth]{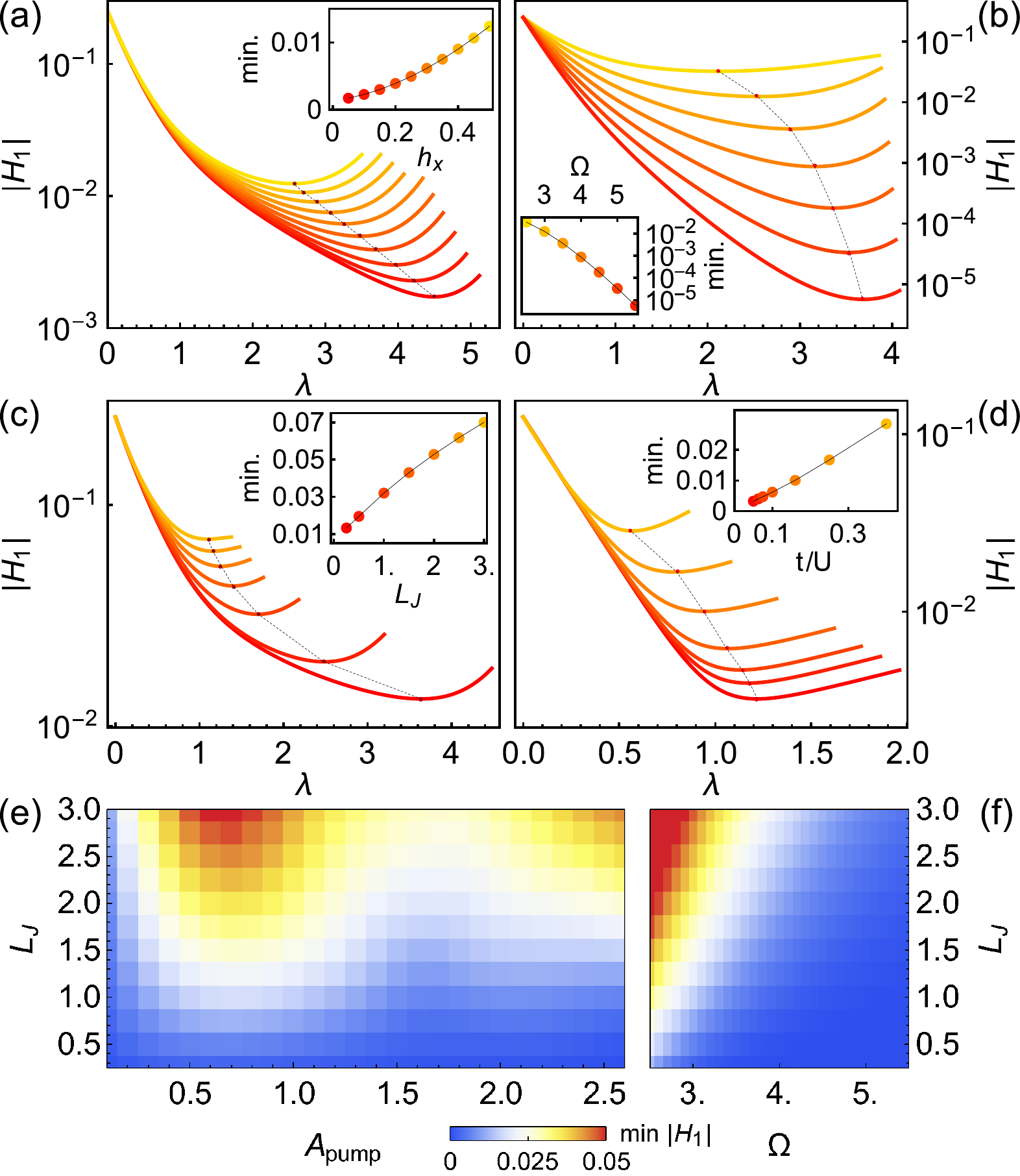}
	\caption{Matrix product operator representation of Floquet flow renormalization and determination of prethermal minima directly in the thermodynamic limit, depicted for the driven exponential Ising model [Eq. (\ref{eq:exponentialIsingModel})] as a function of (a) parallel field $h_x$, (b) frequency, and (c) of the range of Ising exchange $L_J$ couplings. (d) depicts the flow renormalization for the almost resonantly-driven Hubbard model, as a function of $t/U$. Matrix product operators permit an elegant representation of the prethermal regime and loss of locality at finite flow time even for strong driving. Results for the prethermal minima of time-dependent couplings for the exponentially-local Ising model are depicted in (e) as a function of drive strength and range of bare exchange interactions, revealing the appearance of two lobes of fast heating, separated by a longer-lived prethermal plateau for intermediate pump strengths. Conversely, the canonical high-frequency regime is depicted in (f).  }  \label{fig:MPOflows}
\end{figure}

These operator strings constitute an ideal representation of the Floquet renormalization flow in the vicinity of the prethermal fixed points. They efficiently model the growth of subdominant longer-ranged multi-spin interactions, and hence permit a systematic determination of the minima of the flow, to obtain the associated time scales and heating rates directly in the thermodynamic limit. To illustrate this, we study two examples of exponentially-local Hamiltonians for fermions and spins. First, consider a straightforward extension of the Ising model
\begin{align}
	\Ham &= -J \sum_{i < j} e^{-\frac{|i-j-1|}{L_J}} \hat{S}_i^x \hat{S}_j^y  \notag\\
	    &+ \sum_i \left[ h_x \hat{S}_i^x + \left( h_z + A \cos(\Omega t) \right) \hat{S}_i^z \right]  \label{eq:exponentialIsingModel}
\end{align}
to Ising interactions in equilibrium that decay exponentially with distance, parameterized by a length scale $L_J$, and again with the drive entering as a harmonic modulation of the transverse field. An MPO representation follows straightforwardly from Eq. (\ref{eq:exponentialMPO}) via choice $\hat{A}_i = \hat{\mathbf{1}}$. Second, a one-dimensional fermionic Hubbard model
\begin{align}
	\Ham = t_h \sum_{i < j, \sigma} e^{-\frac{|i-j-1|}{L}} \CD{i\sigma} \C{j\sigma} + \left( U + A \cos(\Omega t) \right) \sum_i \ND{i\uparrow} \ND{i\downarrow}
\end{align}
with a periodically-modulated local Coulomb repulsion $U$ \cite{peronaci18} is extended with electron hopping terms that decay exponentially with distance.

The coupled flow equations (\ref{eq:flowHarmonic}) are integrated using an explicit Runge-Kutta stepper for infinite MPO representations of the time-independent $\Ham_0$ and time-dependent $\Ham_1$ component of $\Ham(t)$. MPOs admit addition, scaling and the evaluation of commutators \cite{schollwock11}, each of which again returns an exponentially-local MPO, albeit at potentially extended range. Compression is implemented using a recently-developed algorithm that extends to the thermodynamic limit \cite{parker20}, and convergence is achieved via varying both the flow time step size and bond dimension.

Fig. \ref{fig:MPOflows} depicts the MPO Floquet flow renormalization for exponential Ising and Hubbard models in the thermodynamic limit, simulated for $A=0.5, h_z=0.7, h_x=0.4, \Omega=3$ in units of $J=1$, and $\Omega = 0.975 U$, respectively. The flow progressively eliminates the magnitude of the time-dependent couplings until reaching a minimal value of $\lVert \Ham_1(\lambda) \rVert$ in the vicinity of the prethermal fixed point. As the flow is continued, relevant perturbations push the system away from the prethermal fixed point and towards the thermal fixed point.  This coincides with a loss of locality of the flowing Hamiltonian, leading to a divergence of the MPO bond dimension as well as a breakdown of the iterative MPO compression algorithm which relies on maintaining locality of the Hamiltonian. Crucially however, the prethermal minimum can be resolved accurately, permitting a systematic study of the prethermal regime, time scales and the onset of thermalization.

\begin{figure*}[t]
	\centering
	\includegraphics[width=12cm]{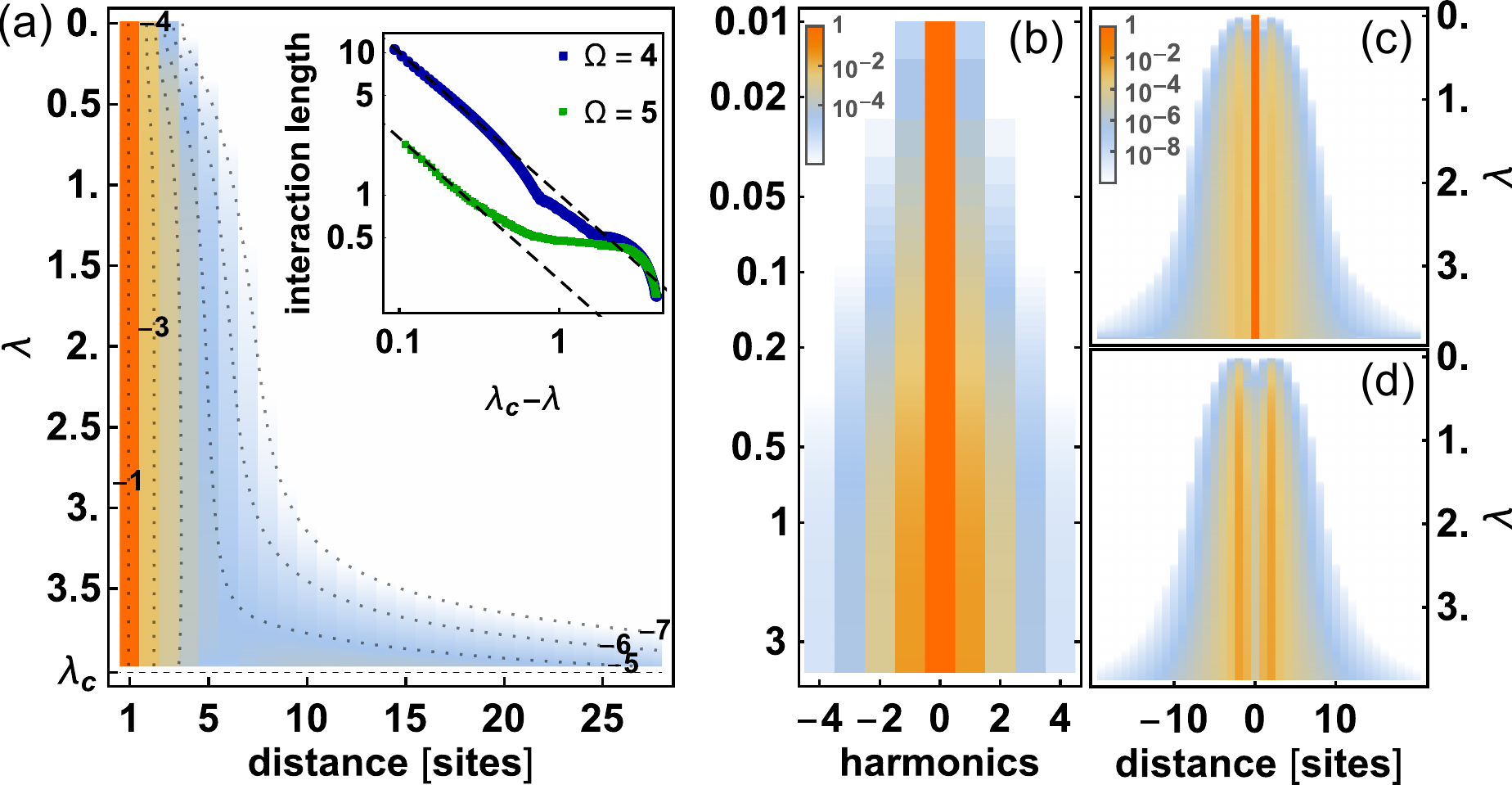}
	\caption{(a) Spreading of the Hamiltonian spin-exchange range as a function of $\lambda$ for an exponential Ising exchange model with $\Omega = 4$. $\Ham_0(\lambda)$ remains exponentially local for small $\lambda$. Beyond the pre-thermal fixed point, flow encounters a power law divergence of the interaction range at a critical $\lambda_c$ at which locality breaks down, as depicted in the inset for the leading exponential behavior for $\Omega = 4, 5$. Flow renormalization of a static local observable $\hat{S}_z$ additionally introduces a periodic time dependence, shown in (b) via norms of the harmonic components. The harmonic weights saturate quickly; conversely (c) and (d) depict spreading of the static and lowest-harmonic components, analogous to the loss of locality of $\Ham_0(\lambda)$.}  \label{fig:operatorSpreading}
\end{figure*}

At large drive frequencies, the minima of the magnitude of time-dependent couplings $\lVert \Ham_1(\lambda) \rVert$ scale exponentially as a function of increasing drive frequency, depicted in Fig. \ref{fig:MPOflows}(b) for the driven Ising model. This corresponds to an exponentially long-lived prethermal regime with time scale $t_{\textrm{eff}} \sim e^{\Omega/\Omega_0}$, which agrees with well-established rigorous bounds for time scales and heating rates the high-frequency regime \cite{abanin15,mori16,kuwahara16,ho17,abanin17a,abanin17b,mori18,ho18}.

Crucially, the flow renormalization procedure goes beyond bounds on the asymptotic behavior and instead permits a direct calculation of the parameter dependence of the prethermal regime and associated time scales. To illustrate this, Fig. \ref{fig:MPOflows}(a) and (c) depict the renormalization flow at fixed frequency as a function of the integrability-breaking parallel field $h_x$ and the decay length $L_J$ of exponentially-ranged Ising interactions, respectively. As the parallel field $h_x$ is reduced, the flow progressively approaches the integrable fixed point as expected until reaching a minimum finite magnitude of the time dependence (and corresponding finite prethermal time scale) due to residual integrability breaking via longer-ranged Ising interactions. Similarly, as the range of Ising interactions $L_J$ is increased, the minimum of $\lVert \Ham_1(\lambda) \rVert$ increases in a sublinear manner with a corresponding shrinking of the prethermal regime. An analogous picture emerges for the driven Hubbard model for driving close to resonance $U$, with the prethermal regime extending linearly as a function of $U/t$. These calculations hence permit a microscopic determination both of the parametric dependences of prethermal time scales (given by the inverse of the minimal value of $\lVert \Ham_1(\lambda) \rVert$, shown in the insets of \ref{fig:MPOflows}) and the effective prethermal Floquet Hamiltonian that governs the intermediate-time behavior.

As the MPO Floquet renormalization flow is inherently non-perturbative, this procedure readily extends to the strong drive regime that lies beyond the reach of perturbative expansions. Fig. \ref{fig:MPOflows}(e) displays the parametric dependence of the prethermal minimum of $\lVert \Ham_1(\lambda) \rVert$ as a function of pump strength $A$ and interaction range $L_J$. Two distinct lobes of rapid heating develop; as the pump strength is progressively increased, remarkably one finds an intermediate parameter regime with suppressed heating and an extended prethermal time scale.

\section{Operator Spreading}
Having established prethermal regimes and their associated time scales from the flow time dependence of the time-dependent couplings, a natural follow-up question concerns the nature of the renormalized effective Hamiltonian near the prethermal minima. Fig. \ref{fig:operatorSpreading}(a) depicts the spreading of the effective interaction range of $\Ham_0(\lambda)$ with flow time $\lambda$ for the driven exponential Ising model (\ref{eq:exponentialIsingModel}) with $h_x = 0.5, h_z = 0.7, L_J = 0.2, A = 0.5$, illustrated by example of the magnitude of generated long-ranged exchange couplings for a set of operator strings of the form $\hat{S}_i^x \hat{P}^\uparrow_{i+1} \hat{P}^\uparrow_{i+2} \cdots \hat{P}^\uparrow_{i+N-1} \hat{S}_{i+N}^x$ with $\hat{P}^\uparrow_i = \tfrac{1}{2} \hat{\mathbf{1}}_i + \hat{S}_i^z$. In equilibrium with almost-vanishing bare interaction range $L_J=0.2$, these rapidly decay exponentially $\sim e^{-N/L_J}$. The Hamiltonian spreads weakly as the theory flows towards the prethermal fixed point, with a slow increase of the exponentially-decaying tail of longer-ranged interactions.

Conversely, the range of the Hamiltonian increases rapidly as the flow continues past the prethermal minimum of the time-dependent couplings. To assess the rate of loss of locality, the tails of generated longer-ranged couplings can be fit via an exponential $e^{-N/\zeta(\lambda)}$. For the same choice of longer-ranged interaction as above, the extracted length scale $\zeta(\lambda)$ is depicted in the inset of Fig. \ref{fig:operatorSpreading}(a). Its behavior at longer flow times is remains bounded at short flow times near the prethermal minimum, and is well-described by a power-law divergence $\zeta \sim (\lambda_c - \lambda)^{-1}$ that signifies a loss of locality at a finite flow time $\lambda_c$, which can be extracted from the fit and corresponds to the onset of heating.

As Hamiltonian flows as governed by the flow renormalization equations (\ref{eq:flowGeneric}), physical observables must similarly transform under the flow. While the flow generator ensures that a harmonically-driven Hamiltonian does not generate higher harmonics of the drive under the flow, no such constraint applies to observables. Instead, the flow renormalization of a generic bare operator $\hat{O}$ will generate a series of harmonics
\begin{align}
	\hat{O}(\lambda, t) = \sum_m e^{i m \Omega t}~ \hat{O}_m(\lambda)
\end{align}
in the rotating frame of the transformation, which are similarly governed by a set of flow equations
\begin{align}
	\frac{\partial}{\partial \lambda} \hat{O}_m(\lambda) = \left[ \Ham_1(\lambda), ~ \hat{O}_{m-1}(\lambda) \right] - \left[ \Ham_1^\dag(\lambda), ~ \hat{O}_{m+1}(\lambda) \right]
\end{align}
with $\hat{O}_{m=0}(0) = \hat{O}$, $\hat{O}_{m \neq 0}(0) = 0$.

In principle, the flow renormalization of a local observable therefore entails a spreading of its support both in real space and in the space of periodic functions. Fig. \ref{fig:operatorSpreading}(b) depicts the early onset of higher-order harmonic components for a local $\hat{S}^z$ operator placed at the center of a 61-site chain, for the exponential Ising model of Eq. (\ref{eq:exponentialIsingModel}). As the Hamiltonian flows towards the prethermal minimum, the observable rapidly spreads to include a sizeable amplitude in its first harmonic component; as the flow continues however, the time-dependent components of the renormalized observable remain confined to the lowest harmonics.

Conversely, Fig. \ref{fig:operatorSpreading}(c) and (d) depict the real-space spreading of the static ($\hat{S}^z_{m=0}$) and first harmonic ($\hat{S}^z_{m=1}$) component of the observable under the flow. By virtue of exponentially-decaying spin interactions of the renormalized drive term $\Ham_1(\lambda)$, the observable acquires an exponential tail, however remains well-localized at the center of the chain upon approaching the prethermal fixed point. As the prethermal minimum is passed, the observable spreads in analogy to the Hamiltonian and locality is lost.

\section{Discussion and Conclusions}

This work introduced a flow renormalization approach for driven quantum systems, which permits studying emergent prethermal dynamical regimes and associated time scales directly from the fixed point structure of the flow, which admits an elegant representation of relevant perturbations in terms of matrix product operators directly in the thermodynamic limit. Fundamentally, this work advocates the study of operator flow renormalization as a tool to understand and characterize dynamical regimes and their associated time scales without necessitating explicit calculations of the real-time dynamics of a driven many-body system; however, an interesting ramification is an efficient scheme to study real-time dynamics on prethermal time scales. Typically, real-time simulations of interacting driven quantum systems are constrained to short times without severe approximations; for instance, matrix product state formulations \cite{schollwock11,paeckel19} suffer from bond dimensions that can grow exponentially in time, unless the drive frequency is very high with respect to the local energy scales of the system \cite{kennes18}. However, entanglement growth is a basis-dependent quantity. Therefore, a straightforward application of the presented results is to utilize the \textit{renormalized} MPO representation of the Hamiltonian $\Ham(\lambda_{\textrm{pre}}, t)$ at the prethermal minimum as an efficient starting point to simulate the real-time evolution. While time evolution in the bare Hamiltonian might exhibit substantial growth of entanglement, the time-dependence of the Hamiltonian of the renormalized theory is minimal at the prethermal minimum, entailing slow energy absorption in the rotating frame and substantially slower growth of entanglement for the evolution of quantum states initially prepared close to the extremal eigenstates of the renormalized time-independent Hamiltonian. An intriguing prospect is the possibility to formulate efficient descriptions of pump-probe spectroscopies of prethermal Floquet states of matter, to guide experiments.

An interesting follow-up question concerns the role of dissipation in stabilizing dynamical regimes in driven systems. Here, the infinite-time limit is well-understood to host non-trivial steady states and phase transitions \cite{sieberer13,eissing16,mathey18} with the energy influx from the pump balanced by dissipation into the environment. However, transient behavior and intermediate-time dynamical regimes in strongly-interacting open systems remain poorly understood; for instance, weak coupling to a bath could entail a competition between the time scale set by the inverse of the dissipation rate and any intrinsic prethermal time scales of the interacting system, suggesting fruitful applications of Floquet flow renormalization methods to driven dissipative superoperators.

Finally, multi-color drives with incommensurate frequencies constitute a natural extension beyond the discrete time periodicity of Floquet engineering \cite{dumitrescu18,peng18,zhao19}. Here, recent results for high-frequency drives established the stability of long-lived prethermal quasi-periodic dynamics \cite{else20}, suggesting routes towards the realization of novel time quasi-crystalline phases. A formalism to understand heating and quasi-stationary prethermal regimes of generic driven systems and their parametric dependence in generic settings hence remains a pertinent topic for future research.

This work was supported by the Flatiron Institute, and by a startup grant from the University of Pennsylvania. The Flatiron Institute is a division of the Simons Foundation.

\bibliography{floquetRG3}

\clearpage
\appendix

\section{Frequency-space representation of generic two-time evolution operators}
\label{app:TimeEvolutionOperator}

The main text employs a frequency-domain formulation of the two-time evolution operator [Eq. (\ref{eq:UT})], as a tool to analyze the real-time behavior of the renormalized Hamiltonian. Given a generic periodically-driven Hamiltonian
\begin{align}
	\Ham(t) = \sum_m e^{im\Omega t} \Ham_m
\end{align}
with a formal complete basis of exact Floquet many-body eigenstates
\begin{align}
	\ket{\Psi_n(t)} = e^{-i\E_n t} \sum_m e^{im\Omega t} \ket{\Phi_{nm}}
\end{align}
with Floquet quasi-energy $\E_n$, the time-ordered time-evolution operator can be exactly expressed as
\begin{align}
	\hat{U}(t,t') &= \hat{\mathcal{T}} e^{-i \int_{t'}^{t} d\tau \Ham(\tau)} = \sum_n \ketbra{\Psi_n(t)}{\Psi_n(t')}
\end{align}
As utilized in the main text, the time ordering can be conveniently circumvented by going to a frequency-domain (Sambe) representation of $\Ham(t)$ which reads
\begin{align}
	\HamF = \sum_{mM} \left[ \Ham_M + M\Omega ~\delta_{M,0} \right] \otimes \ketbra{m+M}{m}
\end{align}
where $m$ indexes frequency-shifted Floquet ``side bands'' that physically identify with, for instance, the number of absorbed or emitted photons in the semi-classical limit of light-matter interaction with an external optical drive.

A full eigenbasis of the frequency-domain Hamiltonian can be labelled as
\begin{align}
	\HamF \ket{\phi_n^{M}} = (\E_n + M\Omega) \ket{\phi_n^{M}}
\end{align}
and appears to naively expand the Hilbert space. This however merely represents the residual Floquet gauge ambiguity, as a shift of the Floquet index of any eigenstate
\begin{align}
	\left(\sum_m \ketbra{m+M'}{m} \right) \ket{\phi_n^{M}} = \ket{\phi_n^{M+M'}}
\end{align}
generates a new eigenstate with an energy shift $M'\Omega$, and hence identifies the same physical state (the physical quasi-energy $\E_n$ is defined modulo $\Omega$). The eigenbasis can now be used to define $\ket{\Phi_{nm}} = \braket{m}{\phi_n^{M}}$. Notably, while the dimension of the expanded Hilbert space of $\HamF$ is in principle infinite, all eigenfunctions $\ket{\phi}$ are exponentially-localized in the space periodic functions around a mean value of $m = \braket{m}{\phi}$. To form a complete basis of Floquet states, it is therefore sufficient to truncate the number of side bands such that $N$ states of $\HamF$ near the center of the spectrum are sufficiently well-converged, with $N$ the dimension of the original Hilbert space of $\Ham(t)$.

Now, rewrite the time-evolution operator as
\begin{align}
	\hat{U}(t,t') &= \sum_n \ketbra{\Psi_n(t)}{\Psi_n(t')} \notag\\
	&= \sum_{nmm'} \ketbra{\Phi_{nm}}{\Phi_{nm'}} e^{-i\E_n (t-t')} e^{i\Omega (mt-m't')} \notag\\
	&= \sum_{nmm'} \braket{m}{\psi_n^{0}} \braket{\psi_n^{0}}{m'} e^{-i\E_n (t-t')} e^{i\Omega (mt-m't')}
\end{align}
Using
\begin{align}
	\braket{m}{\psi_n^{M}} &= \sum_{m'} \braket{m}{m'-M'} \braket{m'}{\phi_{n}^{M+M'}} \notag\\
	&= \braket{m+M'}{\phi_n^{M+M'}}
\end{align}
and the formal matrix exponential of the Sambe Hamiltonian $\HamF$
\begin{align}
	e^{i\HamF t} &= \sum_{nM} \ketbra{\phi_n^{M}}{\phi_n^{M}} e^{-i ( \E_n - M \Omega) t}
\end{align}
one arrives at
\begin{align}
	\hat{U}(t,t') 
	&= \sum_{nmM} \braket{0}{\psi_n^{M}} \braket{\psi_n^{M}}{m} e^{-i (\E_n + M\Omega) (t-t')} e^{-i\Omega m t'} \notag\\
	&= \sum_m \bra{0} e^{i\HamF (t-t')} \ket{m} e^{-i\Omega mt'}
\end{align}
which is the result quoted in the main text. Notably, the rearrangement of the projection onto Floquet harmonics is in principle arbitrary; a useful conjugate expression is
\begin{align}
	\hat{U}(t,t') &= \sum_m \bra{m} e^{i\HamF (t-t')} \ket{0} e^{+i\Omega mt}
\end{align}

\section{Distance metric of the effective time-evolution operator}
\label{app:DistanceMetric}

To compare the behavior of the Floquet renormalization flow to real-time evolution, the main text introduces a distance metric $d(\lambda,t)$ [Eq. \ref{eq:distanceMetric}] that parameterizes how well the time-independent part of the \textit{renormalized} Hamiltonian reproduces the exact real-time evolution. $d(\lambda,t)$ describes the operator distance between the exact ($\hat{U}$) and effective ($\hat{U}_{\textrm{eff}} = e^{i\Ham_0(\lambda) t}$) time-evolution operators, with the starting time averaged over a single drive period to integrate out the residual gauge freedom of Floquet theory. Utilizing the Sambe space representation of the exact time-evolution operator [Eq. (\ref{eq:SambePropagator})], the distance metric can be recast to
\begin{align}
	d^2(\lambda, t) &= \int_0^T \frac{dt_0}{2T} \left\lVert \hat{U}_{\textrm{eff}}(\lambda, t_0 + t, t_0) - \hat{U}(t_0 + t, t_0) \right\rVert^2 \notag\\
		= 1 -&~ \int_0^T \frac{dt_0}{T \tr\{1\}} \textrm{Re} \tr \left\{ \hat{U}^\dag_{\textrm{eff}}(\lambda,t_0+t, t_0) \hat{U}(t_0+t,t_0) \right\} \notag\\
		= 1 -&~ \textrm{Re} \sum_m \int\displaylimits_0^T \frac{dt_0}{T \tr\{1\}} e^{-i m \Omega t_0} \tr \left\{ e^{-i \Ham_0(\lambda) t} \bra{0} e^{-i \HamF(\lambda) t} \ket{m} \right\} \notag\\
		= 1 -&~ \textrm{Re} \tr\left\{ e^{i \HamF_0(\lambda) t} e^{-i \HamF(\lambda) t} \rho_\infty \right\} / \tr\{ \rho_\infty \}
\end{align}
with $\rho_\infty = \mathbf{1} \otimes \ketbra{0}{0}$, and we used a normalized Frobenius norm $\lVert \hat{A} \rVert^2 \equiv \tr(\hat{A}^\dag \hat{A}) / \tr(1)$. The above expression permits a straightforward calculation of the distance metric via truncation of the Sambe space Hamiltonian to only a finite number of harmonics, and without requiring a calculation of the time-ordered time-evolution operator.

To derive the asymptotic behavior of $d(\lambda, t)$, one may use the fact that for times $t = nT$ that are an integer multiple $n$ of the pump period $T$, the Sambe space representation of the propagator is invariant under shifts of the Floquet index:
\begin{align}
	\left[ e^{i \HamF n T},~ \sum_m \ketbra{m+1}{m} \right] = 0
\end{align}
After replacing the Floquet infinite-temperature ensemble by
\begin{align}
	\rho_{\infty} = \mathbf{1} \otimes \frac{1}{2M+1} \sum_{m=-M}^{M} \ketbra{m}{m}
\end{align}
and taking the limit $M \to \infty$, this permits formally replacing the distance metric by its Sambe space representation and reexpressing the trace as an operator norm in Sambe space
\begin{align}
	d(\lambda, nT) &= \left\lVert e^{i \HamF(\lambda) nT} - e^{i \HamF_0(\lambda) nT} \right\rVert_S
\end{align}
where the norm $\lVert \cdot \rVert_S$ is now formally defined to be normalized in Sambe space
\begin{align}
	\lVert \hat{\mathcal{A}} \rVert_S^2 = \tr( \hat{\mathcal{A}}^\dag \hat{\mathcal{A}} ) ~/~ \tr\{ \mathbf{1} \otimes \sum_m \ketbra{m}{m} \}
\end{align}
Armed with a submultiplicative and unitarily equivalent norm of static unitary operators, it is now straightforward to bound the asymptotic behavior. Using an inequality for the difference of squares of unitary operators $\hat{U}$, $\hat{V}$
\begin{align}
	\left\lVert U^2 - V^2 \right\rVert &= \left\lVert U (U - V) + V (U - V) \right\rVert \notag\\
		&\leq \left\lVert U (U - V) \right\rVert + \left\lVert V (U - V) \right\rVert =  2 \left\lVert U - V \right\rVert
\end{align}
one finds by recursion that
\begin{align}
	d(\lambda, nT) &= \lim_{L \to \infty} \left\lVert \left(e^{i \HamF nT / L}\right)^L - \left(e^{i \HamF_0 nT / L}\right)^L \right\rVert_S \notag\\
		&\leq \lim_{L \to \infty} L \left\lVert e^{i \HamF nT / L} - e^{i \HamF_0 nT / L}  \right\rVert_S \notag\\
		&\leq \lim_{L \to \infty} L \left[ \left| \frac{n T}{L} \right| \left\lVert \HamF - \HamF_0 \right\rVert_S + \mathcal{O}(L^{-2}) \right] \notag\\
		&\leq n T \left\lVert \sum_m \sum_{M>0} \left( \Ham_M(\lambda) \otimes \ketbra{m+M}{m} + \hc \right) \right\rVert_S \notag\\
		&\leq 2 n T ~\left\lVert \sum_{M>0} \Ham_M(\lambda) \right\rVert
\end{align}
which recovers expression (\ref{eq:propagatorDistanceBound}) of the main text.

\section{Almost-conservation of the time-independent Hamiltonian near prethermal fixed points}
\label{app:ConservedH0}

Similarly, for the validity of $\Ham_0(\lambda)$ as an effectively-conserved quasi-energy operator, we have
\begin{align}
	C^2&(\lambda,t) = \int_0^T \frac{dt_0}{T} \left\lVert \hat{U}(t_0 + t, t_0) \Ham_0(\lambda) \hat{U}(t_0, t_0 + t) - \Ham_0(\lambda) \right\rVert^2 \notag\\
	    &= 2\left\lVert \Ham_0(\lambda) \right\rVert^2 - 2\textrm{Re} \int_0^T \frac{dt_0}{T} \times \notag\\
	    &~~~~~~~~\times \tr \left\{ \hat{U}(t_0 + t, t_0) \Ham_0(\lambda) \hat{U}(t_0, t_0 + t) \Ham_0(\lambda) \right\}  \notag\\
	    &= 2\left\lVert \Ham_0(\lambda) \right\rVert^2 - 2\textrm{Re} \sum_{mm'} \int_0^T \frac{dt_0~ e^{i(m-m')\Omega t_0}}{T} \times \notag\\
	    &~~~~~~~~\times \tr \left\{ \bra{0} e^{i\HamF t} \ket{m} \Ham_0(\lambda) \bra{m'} e^{-i\HamF t} \ket{0} \Ham_0(\lambda) \right\} \notag\\
	    &= 2\left\lVert \Ham_0(\lambda) \right\rVert^2 - 2\textrm{Re} \tr \left\{ \bra{0} e^{i\HamF t} \tilde{\mathcal{H}}_0(\lambda) e^{-i\HamF t} \tilde{\mathcal{H}}_0(\lambda) \ket{0} \right\} \notag\\
	    &= \left\lVert e^{i\HamF t} \tilde{\mathcal{H}}_0(\lambda) e^{-i\HamF t} - \tilde{\mathcal{H}}_0(\lambda)  \right\rVert_S^2
\end{align}
Using
\begin{align}
	\lVert U U A U^{-1} U^{-1} - A \rVert &= \lVert U (U A U^{-1} - A) U^{-1} ~+ \notag\\
	    &~~~~~~~~~~~+~ (U A U^{-1} - A) \rVert \notag\\
		&\leq 2 \left\lVert U A U^{-1} - A \right\rVert
\end{align}
one obtains a bound
\begin{align}
	C(\lambda,t) &= \lim_{L \to \infty} \left\lVert \left(e^{i\HamF t / L}\right)^L \tilde{\mathcal{H}}_0 \left(e^{-i\HamF t / L}\right)^L - \tilde{\mathcal{H}}_0(\lambda) \right\rVert_S \notag\\
		&\leq \lim_{L \to \infty} L \left\lVert e^{i\HamF t / L} ~\tilde{\mathcal{H}}_0~ e^{-i\HamF t / L} - \tilde{\mathcal{H}}_0 \right\rVert_S \notag\\
		&\leq \lim_{L \to \infty} L \left| \frac{t}{L} \right| \left\lVert \left[ \HamF',~ \tilde{\mathcal{H}}_0 \right] \right\rVert_S \notag\\
		&\leq 2 t \left\lVert \left( \partial_{\lambda} + \Omega \right) \Ham_1  \right\rVert 
\end{align}
which yields the result quoted in the main text.
\clearpage




\end{document}

%% file: macros.tex
\newcommand{\Ham}{\hat{H}}

\newcommand{\ketbra}[2]{\left|#1\middle>\middle<#2\right|}
\newcommand{\braket}[2]{\left<#1\middle|#2\right>}
\newcommand{\braOPket}[3]{\left<#1\middle|#2\middle|#3\right>}
\newcommand{\bra}[1]{\left<#1\right|}
\newcommand{\ket}[1]{\left|#1\right>}
\newcommand{\OPc}[2]{\hat{#1}_{#2}^{\dag}}
\newcommand{\OP}[2]{\hat{#1}_{#2}^{\vphantom{\dag}}}
\newcommand{\CD}[1]{\OPc{c}{#1}}
\newcommand{\C}[1]{\OP{c}{#1}}

\newcommand{\hc}{\textrm{h.c.}}
\newcommand{\E}{\epsilon}

\newcommand{\expect}[1]{\left<#1\right>}